\documentclass{PoS}

\usepackage{amsmath}
\usepackage[utf8]{inputenc}
\usepackage{isotope}

\title{Search for a bound H-dibaryon using local six-quark interpolating operators}

\ShortTitle{Search for a bound H-dibaryon using local six-quark interpolating operators}

\author{\speaker{Jeremy Green},
        Anthony Francis,
        Parikshit Junnarkar,
        Chuan Miao,
        Thomas Rae,
        and Hartmut Wittig
        \\
        Institut für Kernphysik,
        Helmholtz-Institut Mainz,
        and PRISMA Cluster of Excellence,\\
        Johannes Gutenberg-Universität Mainz, D-55099 Mainz, Germany\\
        E-mail:
        \email{green@kph.uni-mainz.de},
        \email{francis@kph.uni-mainz.de},
        \email{junnarka@kph.uni-mainz.de},
        \email{chuan@kph.uni-mainz.de},
        \email{thrae@uni-mainz.de},
        \email{wittig@kph.uni-mainz.de}
      }

      \abstract{We present early results from a lattice QCD study
        seeking a bound $H$-dibaryon using $N_f=2$ flavors of $O(a)$
        improved Wilson fermions and a quenched strange quark. We
        compute a matrix of two-point functions using operators
        consisting of the two independent local products of six
        positive-parity-projected quarks with the appropriate quantum
        numbers, which belong to the singlet and 27-plet irreducible
        representations of flavor SU(3). To expand this basis, we also
        independently vary the quark-field smearing, and apply a new
        scheme to reduce the noise caused by smearing. We then find
        the ground-state mass by solving the generalized eigenvalue
        problem. We show results from ensembles with pion masses
        451~MeV and 1~GeV, and compare with other lattice
        calculations.}

\FullConference{The 32nd International Symposium on Lattice Field Theory\\
		 23--28 June, 2014\\
		 Columbia University New York, NY}

\begin{document}

\section{Introduction}

The $H$-dibaryon is a conjectured six-quark hadron with quark content
$uuddss$ that is a scalar and a flavor
singlet~\cite{Jaffe:1976yi}. Among the experimental constraints on its
existence is the ``Nagara'' event from E373 at
KEK~\cite{Takahashi:2001nm}, which found a
$\isotope[6][\Lambda\Lambda]{He}$ double-hypernucleus with binding
energy $B_{\Lambda\Lambda}=6.91\pm
0.16$~MeV~\cite{Nakazawa:2010zza}. The absence of a strong decay
$\isotope[6][\Lambda\Lambda]{He} \to \isotope[4]{He} + H$ implies
\begin{equation}
m_H > 2m_\Lambda - B_{\Lambda\Lambda}.
\end{equation}

Lattice QCD studies of a possible $H$-dibaryon have been done for
nearly thirty years~\cite{Mackenzie:1985vv}; however, aside from this
work, calculations with dynamical fermions have only been done by two
collaborations: NPLQCD~\cite{Beane:2010hg,Beane:2011zpa,Beane:2012vq}
and HALQCD~\cite{Inoue:2010es,Inoue:2011ai,Inoue:2012jh}. These
calculations have generally found a bound $H$-dibaryon at fairly heavy
quark masses, with a binding energy that becomes smaller when the
quark masses are decreased.

\section{Lattice methodology}

For a set of interpolating operators $\mathcal{O}_i$, we compute a
matrix of zero-momentum two-point functions,
\begin{equation}
  C_{ij}(t) = \sum_x \langle \mathcal{O}^{\vphantom{\dagger}}_i(x,t_0+t)\mathcal{O}_j^\dagger(x_0,t_0)\rangle.
\end{equation}
We then find effective masses in two ways: from its diagonal elements,
\begin{equation}
  m_{\text{eff},i}(t) = \frac{1}{\Delta t}\log\frac{C_{ii}(t)}{C_{ii}(t+\Delta t)},
\end{equation}
and from using the variational method~\cite{Luscher:1990ck} by solving
the generalized eigenvalue problem (GEVP),
\begin{equation}
 C_{ij}(t+\Delta t)v_j(t) = \lambda(t)C_{ij}(t)v_j(t);\quad
m_\text{eff}(t) = \frac{-\log\lambda(t)}{\Delta t}.
\end{equation}
Both of these effective masses will approach the ground-state mass
from above, with exponentially-decaying excited-state contamination;
the variational method serves to eliminate contributions from the
lowest-lying excited states~\cite{Blossier:2009kd}.

We construct interpolating operators using products of six
positive-parity-projected smeared quarks fields, of the form
\begin{equation}\label{eq:6quark}
 [abcdef] = \epsilon^{ijk}\epsilon^{lmn} (b^T_i C\gamma_5P_+ c_j)
(e_l^T C\gamma_5P_+ f_m) (a^T_k C\gamma_5P_+ d_n),
\end{equation}
where $P_+ = (1+\gamma_0)/2$. There are two such local operators that
can couple to an $H$-dibaryon in a calculation with degenerate $u$ and
$d$ quarks~\cite{Donoghue:1986zd,Wetzorke:1999rt}:
\begin{align}
H^\mathbf{1} &= \frac{1}{48}\left([sudsud] - [udusds] - [dudsus]\right)\\
H^\mathbf{27}&= \frac{1}{48\sqrt{3}}\left(3[sudsud] + [udusds] + [dudsus]\right),
\end{align}
which belong to the singlet and 27-plet irreps of flavor SU(3).

The standard smearing procedure uses Wuppertal
smearing~\cite{Gusken:1989qx},
\begin{equation}
\tilde q = (1 + \alpha H)^n q,
\end{equation}
where $H$ is the gauge-covariant spatial hopping term constructed
using spatially APE-smeared gauge links~\cite{Albanese:1987ds};
generically we write the smearing kernel as $S_{ab}(x,y;t)$.  This
tends to introduce noise due to fluctuations in the shape of the
smeared field, especially after many steps of
smearing~\cite{vonHippel:2013yfa}. To first approximation, these
fluctuations affect the overall amplitude of the smeared quark field
and could be reduced by normalizing:
\begin{equation}
\tilde q_{N1}(x,t) = \frac{1}{N(x,t)}\tilde q(x,t), \quad N(x,t) =
\sqrt{\sum_{y,a,b} |S_{ab}(x,y;t)|^2}.
\end{equation}
This is simple to apply to a point source but would be rather
expensive to compute at the sink. Instead, we introduce a variant of
this method, \emph{timeslice-normalized smearing}, with the
normalization factor summed over each timeslice using stochastic
estimation:
\begin{equation}
\tilde q_N(x,t) = \frac{1}{N(t)}\tilde q(x,t), \quad N(t)^2 =
\frac{1}{n_\text{noise}}\sum_{x,y,a,b,i}|S_{ab}(x,y;t)\eta_b^{(i)}(y,t)|^2
\approx \sum_{x,y,a,b} |S_{ab}(x,y;t)|^2,
\end{equation}
where $\eta^{(i)}$ are
noise vector fields with expectation value
$E[\eta^{(i)}\eta^{(j)\dagger}]=\delta^{ij}I$. A key feature of this
prescription is that a correlator with timeslice-normalized smearing
can be obtained from one with ordinary smearing by simple
multiplication of the smearing factors: for a dibaryon operator,
$C_N(t_f,t_i)=(\frac{1}{N(t_i)N(t_f)})^6C(t_f,t_i)$, where this is
applied individually to each sample\footnote{Note that although
  $E[N(t)^2]=\sum_x N(x,t)^2$ does not imply $E[N(t)^{-6}]=(\sum_x
  N(x,t)^2)^{-3}$, with sufficiently many noise sources the difference
  is small, and its symmetry properties are never spoiled. To be
  precise: as long as $t_f\neq t_i$, the expectation value of $C_N$ is
  a two-point function constructed from quark fields $\tilde
  q_N(x,t)=E[N(t)^{-6}]^{1/6}\tilde q(x,t)$, which depend on the
  precise details of the set of noise vectors.}. In practice, we used
160 $Z_4$ noise vectors with color dilution, which has the same
computational cost as smearing ten propagators.

To obtain high statistics, we also make use of the all-mode-averaging
(AMA) technique~\cite{Blum:2012uh,Shintani_poster}, computing many
samples with lower-precision propagator solves and applying a bias
correction using a relatively small number of high-precision solves.

\section{Results}

We performed calculations on two ensembles generated within and
provided to us by the CLS effort, with $N_f=2$ non-perturbatively
clover-improved Wilson fermions, labeled E1 and
E5~\cite{Fritzsch:2012wq}, which have lattice spacing
$a=0.063$~fm~\cite{Capitani:2011fg} and lattice volume $64\times
32^3$. We used a quenched strange quark with hopping parameter
$\kappa_s$ tuned such that the ratio
$(m_K^2-\frac{1}{2}m_\pi^2)/m_\Omega^2$ takes its physical value.
However, on the E1 ensemble we deviated slightly from this tuning to
set $\kappa_s$ equal to the light-quark hopping parameter in order to
have SU(3)-flavor symmetry. The basic information about these two
ensembles is summarized in Table~\ref{tab:ensembles}.

\begin{table}
  \centering
  \begin{tabular}{l|cc|cc|rrr}
    Label & $\kappa_{ud}$ & $\kappa_s$ & $m_\pi$ & $m_\pi L$ & $N_\text{conf}$ & $N_\text{src}$ & $N_\text{samp}$ \\\hline
    E1 & 0.13550 & 0.135500& 1 GeV   & 10  & 168 & 128 & 43008 \\
    E5 & 0.13625 & 0.135546& 451 MeV & 4.6 &1881 &  16 & 60192
  \end{tabular}
  \caption{Ensemble parameters. Note that $N_\text{src}$ indicates the number of low-precision source points used per configuration; in both cases one high-precision source was used. We additionally construct operators using $P_-=(1-\gamma_0)/2$ in place of $P_+$ in Eq.~(\protect\ref{eq:6quark})
 to yield both forward and backward-propagating states from each source and thus two samples per source, using time-reversal symmetry.}
  \label{tab:ensembles}
\end{table}

\begin{figure}
  \centering
  \includegraphics[width=0.49\textwidth]{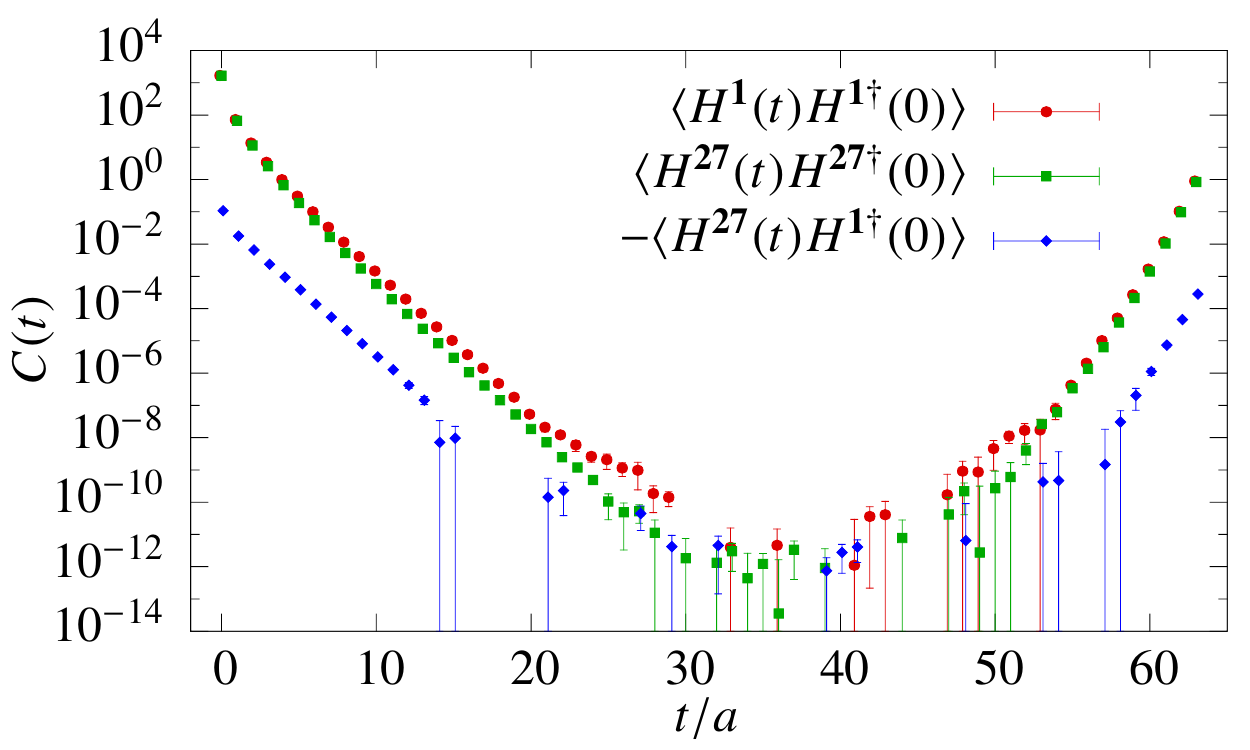}
  \includegraphics[width=0.49\textwidth]{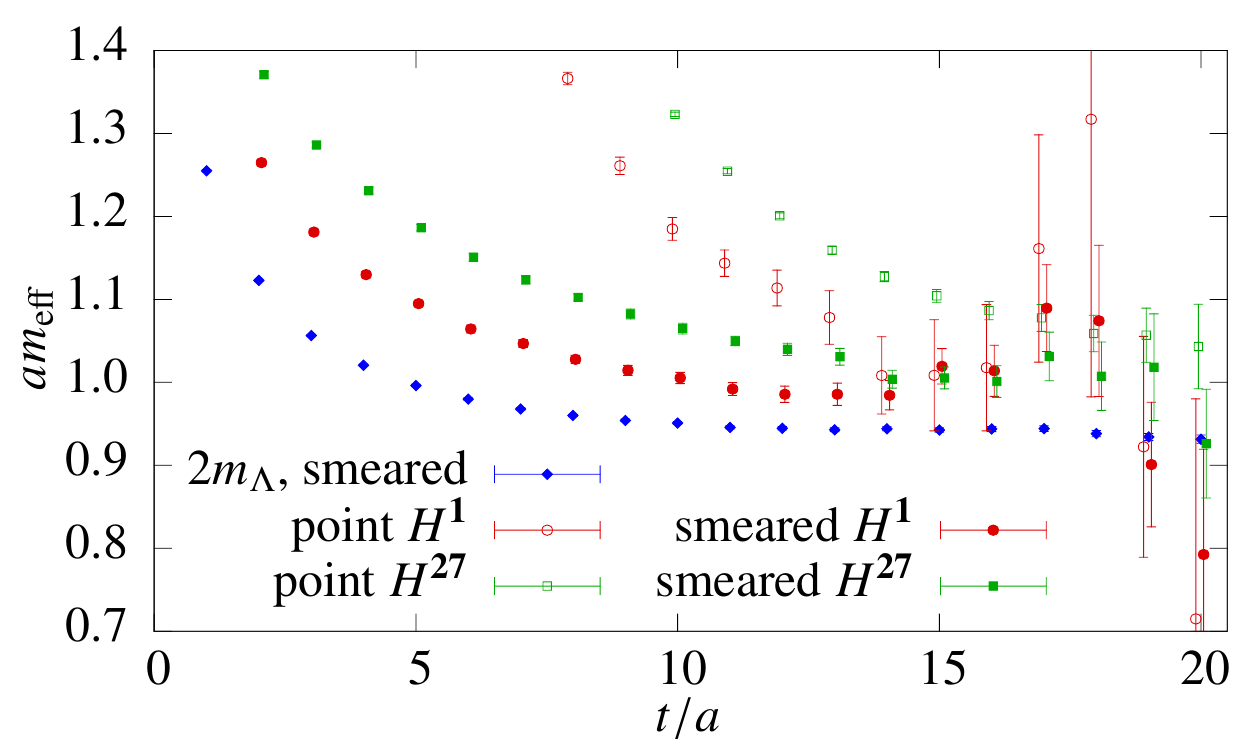}
  \caption{Data from the E5 ensemble. \textbf{Left:} Diagonal and
    off-diagonal two-point functions of the two smeared
    operators. \textbf{Right:} Diagonal effective masses of the four
    operators, along with twice the $\Lambda$ effective mass.}
  \label{fig:E5_c2pt_diagmeff}
\end{figure}

The E5 ensemble has a pion mass of 451~MeV. We computed $H$-dibaryon
two-point functions using both point and smeared ($n=140$,
$\alpha\approx 0.75$) quark fields. Together with the singlet and
27-plet combinations, this yields four interpolating operators. The
two-point functions computed with smeared operators are shown in
Fig.~\ref{fig:E5_c2pt_diagmeff} (left); we find that the
cross-correlator between the singlet and 27-plet operators is
suppressed by 2--3 orders of magnitude, indicating that breaking of
flavor-SU(3) is small.

\begin{figure}
  \centering
  \includegraphics[width=0.49\textwidth]{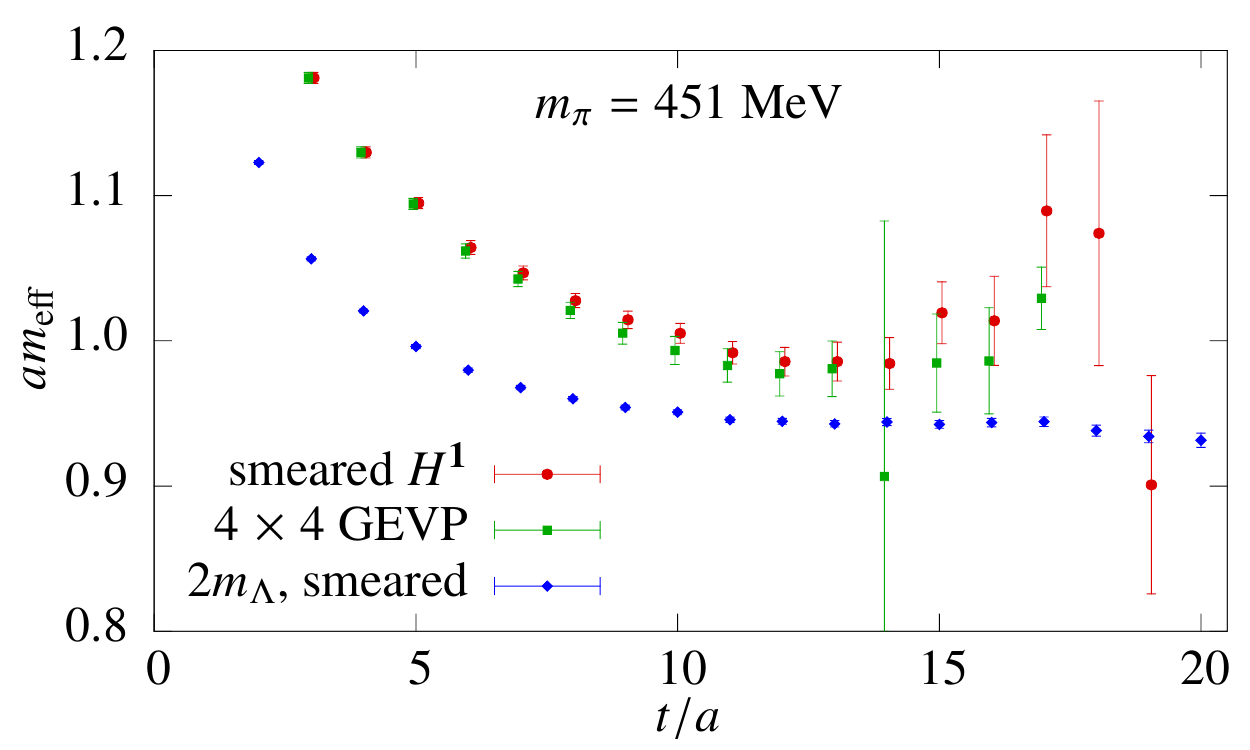}
  \includegraphics[width=0.49\textwidth]{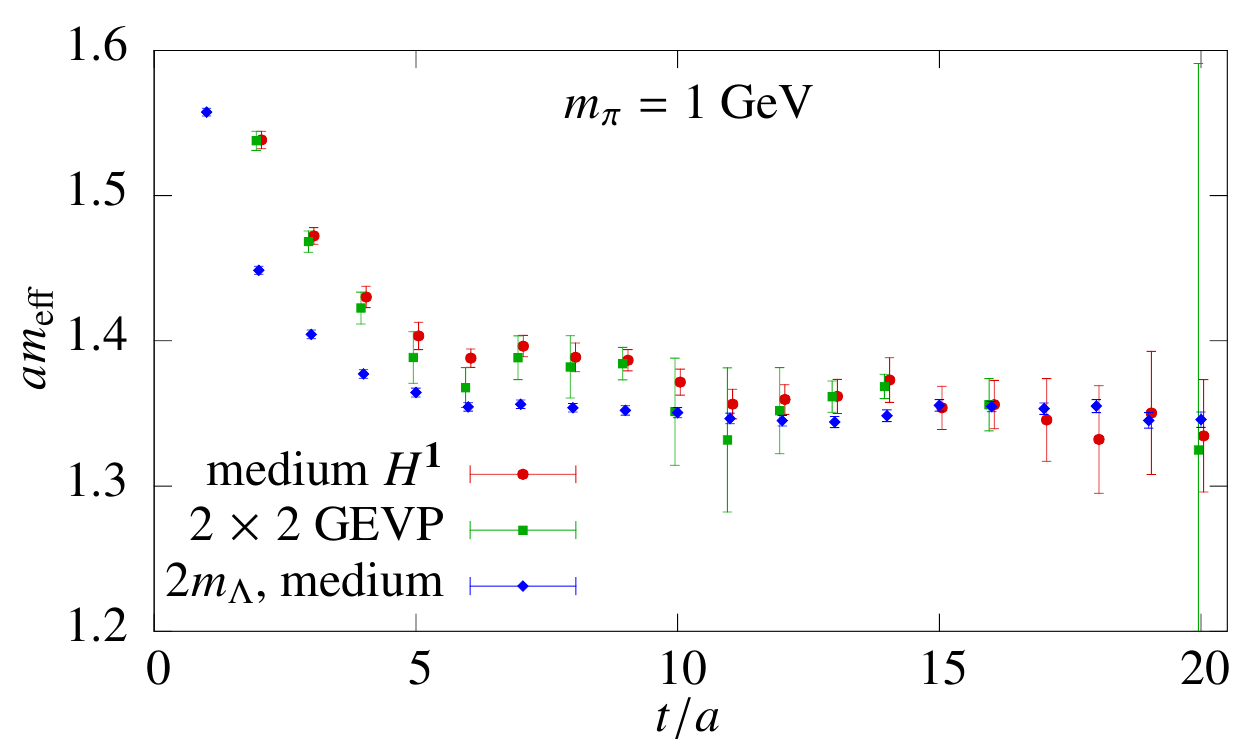} 
  \caption{Effective masses on the two ensembles, in each case
    comparing results from the ``best'' single operator (see text) and
    from the variational method against twice the $\Lambda$ effective
    mass. \textbf{Left:} E5 ensemble. \textbf{Right:} E1 ensemble.}
  \label{fig:E5_E1_GEVP}
\end{figure}

As the lattice spacing is fairly fine, we compute effective masses
with a time step $\Delta t=3a$. For the four diagonal two-point
functions, the effective masses are shown in
Fig.~\ref{fig:E5_c2pt_diagmeff} (right), along with twice the
effective mass of the $\Lambda$ baryon. All four $H$-dibaryon
effective masses remain clearly above the $2m_\Lambda$ threshold until
their signals deteriorate.
As should be expected, smearing reduces the coupling to excited
states. In addition, the effective masses of the flavor-singlet
operators approach the ground state faster than the 27-plet
operators. Therefore, we consider the singlet operator with smearing
as the ``best'' single operator; indeed, it is the first to reach a
possible plateau at around $t/a=12$. For a final analysis on this
ensemble, we use the variational method with the full $4\times 4$
matrix of two-point functions. The effective mass as determined from
the GEVP is shown in Fig.~\ref{fig:E5_E1_GEVP} (left), alongside the
effective mass from the ``best'' single operator and twice the
$\Lambda$ effective mass. The variational approach yields effective
masses that are slightly lower than the ``best'' operator, and thus a
faster approach to a plateau, at the cost of increased statistical
uncertainty. Neither approach shows a clear sign of a plateau below
the $2m_\Lambda$ threshold and we conclude that the data on this
ensemble do not indicate the presence of a bound $H$-dibaryon.

As other calculations have found a more strongly bound state at higher
pion masses, we used the E1 ensemble, which has a pion mass of 1~GeV,
for comparison. Since the light and strange quarks are degenerate,
there is no mixing between the singlet and 27-plet
operators. Therefore, in order to expand the basis of operators in the
singlet channel, we used additional smearings: in addition to the same
smearing used on E5, which we call ``medium'', we added ``narrow'' and
``wide'' with half and twice as many steps, for a total of three
singlet interpolating operators.

\begin{figure}
  \centering
  \includegraphics[width=0.49\textwidth]{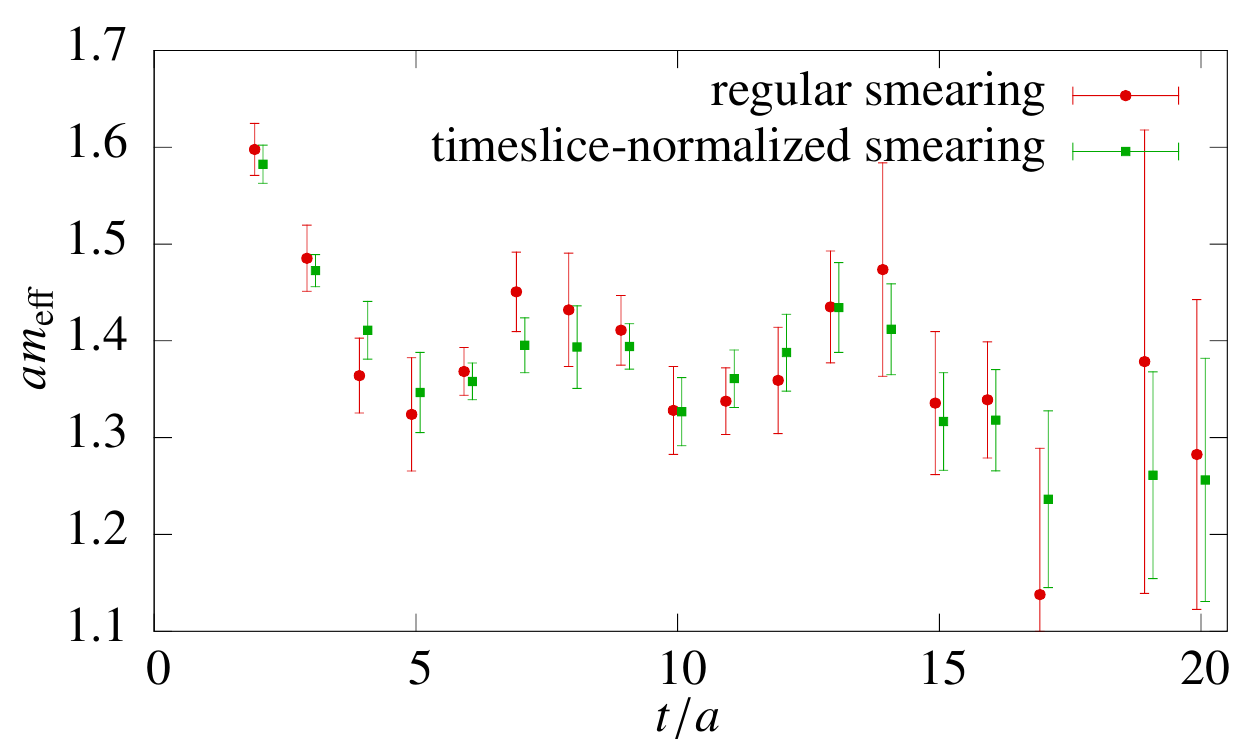}
  \includegraphics[width=0.49\textwidth]{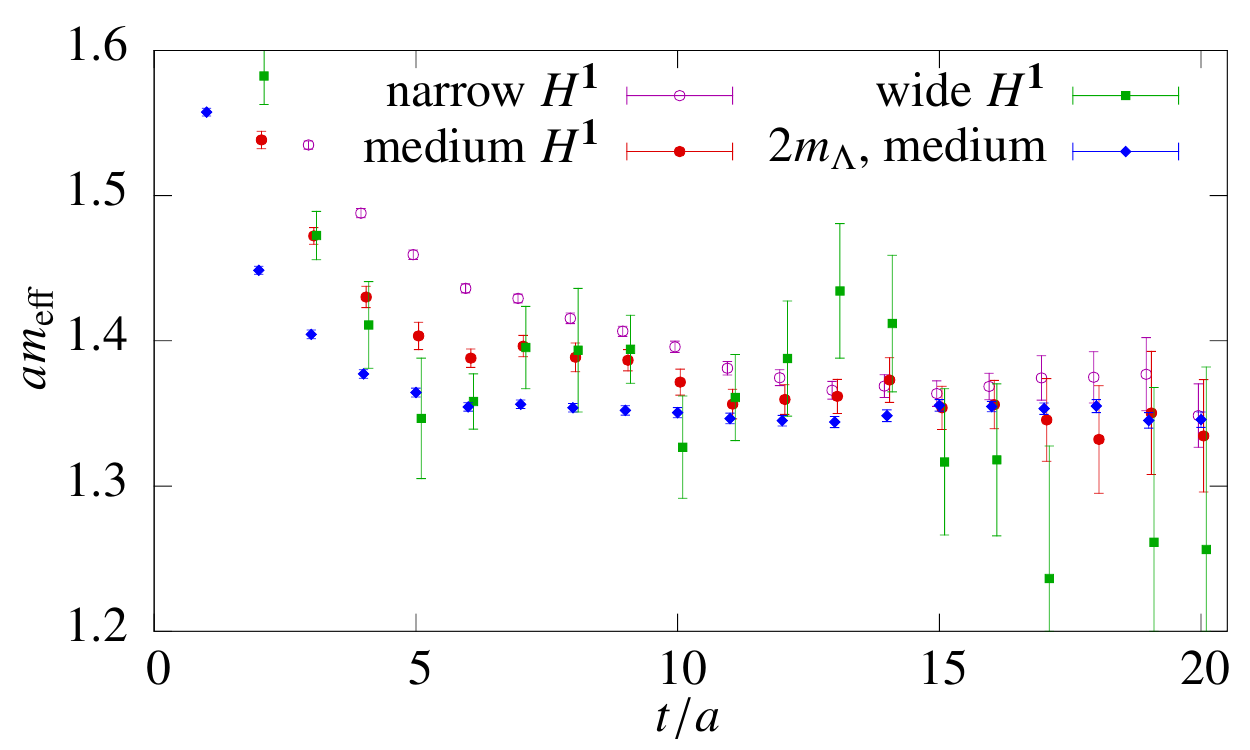}
  \caption{Effective masses from the E1 ensemble. \textbf{Left:}
    Comparison of regular and timeslice-normalized smearing, for the
    ``wide'' singlet operator. \textbf{Right:} Diagonal effective
    masses of the three operators, along with twice the $\Lambda$
    effective mass.}
  \label{fig:E1_smearing_diagmeff}
\end{figure}

The ``wide'' smearing introduces additional noise, and thus it
benefits the most from timeslice-normalized smearing. This is shown in
Fig.~\ref{fig:E1_smearing_diagmeff} (left), where a considerable
reduction in noise can be seen in this operator's effective mass. The
diagonal effective masses for the three singlet operators are shown in
Fig.~\ref{fig:E1_smearing_diagmeff} (right). The ``wide'' operator
approaches a plateau at a similar rate as the ``medium'' operator,
although it is much noisier. The ``narrow'' operator has the least
noise of the three, but it has greater contamination from excited
states. The plateaus reached by the operators are consistent with the
$2m_\Lambda$ threshold; i.e., present statistics are insufficient to
determine if the plateaus lie above or below it.

For the variational analysis on E1, we omit the ``wide'' operator due
to its high level of noise and solve the GEVP for the remaining
$2\times 2$ matrix of two-point functions. The resulting effective
mass is shown in Fig.~\ref{fig:E5_E1_GEVP} (right), alongside the
effective mass from the ``medium'' operator and
$2m_{\text{eff},\Lambda}$. As was the case for E5, the
variational-method result has a slightly faster approach to a plateau,
however it is somewhat noisier than the single operator. There is no
plateau clearly below the $2m_\Lambda$ threshold, and thus we do not
find evidence for a bound $H$-dibaryon on this ensemble; however,
present statistics are insufficient to rule out the possibility.

\section{Discussion and outlook}
The absence of a bound $H$-dibaryon is in contrast with other lattice
QCD calculations using dynamical fermions. In particular, NPLQCD,
computing the spectrum using two-point functions on the lattice, found
a binding energy of 75~MeV using ensembles with three degenerate quark
flavors and pion masses of about 800~MeV~\cite{Beane:2012vq}, which
corresponds to about 0.024 in lattice units on the ``E''
lattices. This is clearly inconsistent with the data on E1, except at
rather large source-sink separations where the statistical uncertainty
is much greater.

The lattice ensembles used in this work and by NPLQCD are somewhat
different. In particular, this work lacks a finite-volume study, which
may be important for multi-baryon states; the volumes used in
Ref.~\cite{Beane:2012vq} were all larger than the single volume used
in this work, however no significant dependence of the $H$-dibaryon
binding energy on the lattice volume was found in that work. In
addition, this work used a significantly finer lattice spacing and a
quenched strange quark; it is possible that these contribute to the
difference in results.

A key difference between this work and that of NPLQCD is the
interpolating operators: whereas we use a symmetric setup with the
same six-quark operators at the source (at a single point) and at the
sink (projected to zero momentum), NPLQCD used an asymmetric setup.  At
the source, they had six-quark operators at a point, and at the sink,
they used a product of two individually momentum-projected three-quark
operators (i.e., two-baryon operators). It may be the case that the
two-baryon operators couple much better to the ground state in this
channel, and that very large source-sink separations would be required
to obtain the ground-state energy using only six-quark operators.

Another related possibility is that a reduced statistical uncertainty
would reveal a plateau below the $2m_\Lambda$ threshold at somewhat
larger source-sink separations.

We plan to address both of the latter possibilities by supplementing
our set of operators with two-baryon operators at the sink, in
addition to increasing statistics. This will allow for a cleaner
comparison with NPLQCD, as well as the qualitative information about
the ground state that can be obtained by studying its relative
couplings to six-quark and two-baryon operators.

\acknowledgments

We thank our colleagues within CLS for sharing the lattice ensembles
used.  This work made use of the ``Wilson'' cluster at the Institute
for Nuclear Physics and the ``Clover'' cluster at the Helmholtz
Institute Mainz, both at the University of Mainz; we thank Christian
Seiwerth and Dalibor Djukanovic for technical support. We are also
grateful for computer time allocated to project HMZ21 on the BG/Q
JUQUEEN computer at NIC, Jülich.

\bibliographystyle{JHEP-2}
\bibliography{dibaryon.bib}

\providecommand{\href}[2]{#2}\begingroup\raggedright\begin{thebibliography}{10}

\bibitem{Jaffe:1976yi}
R.~L. Jaffe, {\it Perhaps a stable dihyperon},  {\em Phys. Rev. Lett.} {\bf 38}
  (1977) 195--198.

\bibitem{Takahashi:2001nm}
H.~Takahashi, J.~K. Ahn, H.~Akikawa, S.~Aoki, K.~Arai {\em et~al.}, {\it
  Observation of a {$\isotope[6][\Lambda\Lambda]{He}$} double hypernucleus},
  {\em Phys. Rev. Lett.} {\bf 87} (2001) 212502.

\bibitem{Nakazawa:2010zza}
{\bf KEK-E176, E373, and J-PARC-E07} Collaboration, K.~Nakazawa, {\it
  {Double-$\Lambda$ hypernuclei via the $\Xi^-$ hyperon capture at rest
  reaction in a hybrid emulsion}},  {\em Nucl. Phys. A} {\bf 835} (2010)
  207--214.

\bibitem{Mackenzie:1985vv}
P.~B. Mackenzie and H.~B. Thacker, {\it Evidence against a stable dibaryon from
  lattice {QCD}},  {\em Phys. Rev. Lett.} {\bf 55} (1985) 2539.

\bibitem{Beane:2010hg}
{\bf NPLQCD} Collaboration, S.~R. Beane {\em et~al.}, {\it Evidence for a bound
  {H}-dibaryon from lattice {QCD}},  {\em Phys. Rev. Lett.} {\bf 106} (2011)
  162001 [\href{http://arXiv.org/abs/1012.3812}{{\tt 1012.3812}}].

\bibitem{Beane:2011zpa}
{\bf NPLQCD} Collaboration, S.~R. Beane, E.~Chang, W.~Detmold, B.~Joó, H.~W.
  Lin {\em et~al.}, {\it Present constraints on the {H}-dibaryon at the
  physical point from lattice {QCD}},  {\em Mod. Phys. Lett. A} {\bf 26} (2011)
  2587--2595 [\href{http://arXiv.org/abs/1103.2821}{{\tt 1103.2821}}].

\bibitem{Beane:2012vq}
{\bf NPLQCD} Collaboration, S.~R. Beane, E.~Chang, S.~D. Cohen, W.~Detmold,
  H.~W. Lin {\em et~al.}, {\it Light nuclei and hypernuclei from quantum
  chromodynamics in the limit of {SU(3)} flavor symmetry},  {\em Phys. Rev. D}
  {\bf 87} (2013) 034506 [\href{http://arXiv.org/abs/1206.5219}{{\tt
  1206.5219}}].

\bibitem{Inoue:2010es}
{\bf HAL QCD} Collaboration, T.~Inoue {\em et~al.}, {\it Bound {H}-dibaryon in
  flavor {SU(3)} limit of lattice {QCD}},  {\em Phys. Rev. Lett.} {\bf 106}
  (2011) 162002 [\href{http://arXiv.org/abs/1012.5928}{{\tt 1012.5928}}].

\bibitem{Inoue:2011ai}
{\bf HAL QCD} Collaboration, T.~Inoue {\em et~al.}, {\it Two-baryon potentials
  and {H}-dibaryon from 3-flavor lattice {QCD} simulations},  {\em Nucl. Phys.
  A} {\bf 881} (2012) 28--43 [\href{http://arXiv.org/abs/1112.5926}{{\tt
  1112.5926}}].

\bibitem{Inoue:2012jh}
{\bf HAL QCD} Collaboration, T.~Inoue, {\it Study of {H}-dibaryon mass in
  lattice {QCD}},  {\em PoS} {\bf LATTICE2012} (2012) 144
  [\href{http://arXiv.org/abs/1212.4230}{{\tt 1212.4230}}].

\bibitem{Luscher:1990ck}
M.~Lüscher and U.~Wolff, {\it How to calculate the elastic scattering matrix
  in two-dimensional quantum field theories by numerical simulation},  {\em
  Nucl. Phys. B} {\bf 339} (1990) 222--252.

\bibitem{Blossier:2009kd}
B.~Blossier, M.~Della~Morte, G.~von Hippel, T.~Mendes and R.~Sommer, {\it {On
  the generalized eigenvalue method for energies and matrix elements in lattice
  field theory}},  {\em JHEP} {\bf 0904} (2009) 094
  [\href{http://arXiv.org/abs/0902.1265}{{\tt 0902.1265}}].

\bibitem{Donoghue:1986zd}
J.~F. Donoghue, E.~Golowich and B.~R. Holstein, {\it Weak decays of the {$H$}
  dibaryon},  {\em Phys. Rev. D} {\bf 34} (1986) 3434.

\bibitem{Wetzorke:1999rt}
I.~Wetzorke, F.~Karsch and E.~Laermann, {\it Further evidence for an unstable
  {H} dibaryon?},  {\em Nucl. Phys. Proc. Suppl.} {\bf 83} (2000) 218--220
  [\href{http://arXiv.org/abs/hep-lat/9909037}{{\tt hep-lat/9909037}}].

\bibitem{Gusken:1989qx}
S.~Güsken, {\it {A Study of smearing techniques for hadron correlation
  functions}},  {\em Nucl. Phys. Proc. Suppl.} {\bf 17} (1990) 361--364.

\bibitem{Albanese:1987ds}
{\bf APE} Collaboration, M.~Albanese {\em et~al.}, {\it Glueball masses and
  string tension in lattice {QCD}},  {\em Phys. Lett. B} {\bf 192} (1987)
  163--169.

\bibitem{vonHippel:2013yfa}
G.~M. von Hippel, B.~Jäger, T.~D. Rae and H.~Wittig, {\it The shape of
  covariantly smeared sources in lattice {QCD}},  {\em JHEP} {\bf 1309} (2013)
  014 [\href{http://arXiv.org/abs/1306.1440}{{\tt 1306.1440}}].

\bibitem{Blum:2012uh}
T.~Blum, T.~Izubuchi and E.~Shintani, {\it {New class of variance-reduction
  techniques using lattice symmetries}},  {\em Phys. Rev. D} {\bf 88} (2013)
  094503 [\href{http://arXiv.org/abs/1208.4349}{{\tt 1208.4349}}].

\bibitem{Shintani_poster}
E.~Shintani, {\it Error reduction with all-mode-averaging in {Wilson} fermion},
   \pos{PoS(LATTICE2014)124}.

\bibitem{Fritzsch:2012wq}
P.~Fritzsch, F.~Knechtli, B.~Leder, M.~Marinkovic, S.~Schaefer {\em et~al.},
  {\it {The strange quark mass and Lambda parameter of two flavor QCD}},  {\em
  Nucl. Phys. B} {\bf 865} (2012) 397--429
  [\href{http://arXiv.org/abs/1205.5380}{{\tt 1205.5380}}].

\bibitem{Capitani:2011fg}
S.~Capitani, M.~Della~Morte, G.~von Hippel, B.~Knippschild and H.~Wittig, {\it
  {Scale setting via the $\Omega$ baryon mass}},  {\em PoS} {\bf LATTICE2011}
  (2011) 145 [\href{http://arXiv.org/abs/1110.6365}{{\tt 1110.6365}}].

\end{thebibliography}\endgroup

\end{document}